\documentclass[twocolumn,showpacs,preprintnumbers,amsmath,amssymb,aps,noshowpacs,pre]{revtex4}

\usepackage{latexsym}
\usepackage{dcolumn}% Align table columns on decimal point
\usepackage{supertabular,multirow}
\usepackage{epsfig}
\usepackage{bm}% bold math
\usepackage{subfigure}
\usepackage{amsmath}
\usepackage{graphicx}
\usepackage{color}

\definecolor{brown}{rgb}{0.59, 0.29, 0.0}
 \definecolor{orange}{RGB}{255,127,0}
\definecolor{brightube}{rgb}{0.82, 0.62, 0.91}

\usepackage{epstopdf}%This line makes .eps figures into .pdf - please comment out if not required.

\definecolor{cream}{RGB}{222,217,201}

\begin{document}

\title{ Dynamics of inert spheres in active suspensions of micro-rotors}

\author{Kyongmin Yeo$^{1,2}$, Enkeleida Lushi$^3$, Petia M. Vlahovska$^3$ }

\affiliation{$^1$IBM T.J. Watson Research Center, Yorktown Heights, NY 10598, USA\\
$^2$Division of Applied Mathematics, Brown University, RI 02912, USA\\
$^3$School of Engineering, Brown University, RI 02912, USA}

\date{\today}

\begin{abstract}
Inert particles suspended in active fluids of self-propelled particles are known to often exhibit enhanced diffusion and novel coherent structures. Here we numerically investigate the dynamical behavior and self-organization in a system consisting of passive and actively rotating spheres. The particles interact through direct collisions and the fluid flows generated as they move. In the absence of passive particles, three states emerge in a binary mixture of spinning spheres depending on particle fraction: a dilute gas-like state where the rotors move chaotically, a phase-separated state where like-rotors move in lanes or vortices, and a jammed state where crystals continuously assemble, melt and move (K. Yeo, E. Lushi, and P. M. Vlahovska, Phys. Rev. Lett. {\bf114}, 188301 (2015)). Passive particles added to the rotor suspension modify the system dynamics and pattern formation: while states identified in the pure active suspension still emerge, they occur at different densities and mixture proportions. The dynamical behavior of the inert  particles is also non-trivially dependent on the system composition.

\end{abstract}

%\pacs{47.57.E-,47.63.mf, 83.10.Tv, 64.75.Xc}% PACS, the Physics and Astronomy

\maketitle

\section{Introduction}

Self-driven (active) particles exhibit collective behavior relevant to understanding  many phenomena in living systems: from  colonies of bacteria and algae,  to microtubule assemblies, to  schools of fish \cite{Marchetti2013}. Self-propelled particles such as swimming bacteria \cite{Dunkel2013} or chemically-propelled colloids \cite{Theurkauff2012} are known to exhibit intriguing collective dynamics and feature macroscopic flows on a scale larger than the individual particles \cite{Sokolov2007, Dunkel2013}, enhanced mixing \cite{Kim2004, Mino2011, Angelani2011, Jepson2013, Kasyap2014, Patteson2015}, as well as size- and shape-dependent diffusion of inert (passive) particles immersed in the active bath \cite{Patteson2015, Cheng2015}.   

Particle diffusion in non-equilibrium systems such as active fluids is a topic of increasing interest \cite{Patteson2015}. The motion of a inert (tracer) particle in the disturbance flow created by an individual micro-swimmer has been considered experimentally \cite{WuLibchaber2000, Leptos2009, Jeaneret2016} and theoretically \cite{LinThiffeault2011, Pushkin2013}. Denser mixtures of passive and self-propelling particles has been extensively studied using simulations \cite{Hinz2014,Li2015,Takatori2015}, especially in the context of activity-induced phase-separation \cite{Fily2012a, Redner2013, Stenhammar2015}.

Unlike self-propelled particles such as bacteria and colloids, spinning particles (``rotors'') and their collective dynamics are far less explored \cite{Snezhko2015b} mostly because there are fewer experimental realizations, e.g., magnetically- driven colloids \cite{Kokot2013, Kokot2015, Snezhko2015} and electrically-driven ``Quincke'' colloids \cite{Bricard2013,Bricard2015}.
Self-organization in rotor suspensions have been analyzed mostly computationally \cite{NguyenKEG14, Spellings2015, Sabrina2015, GotoTanaka2015, YeoLV15}, though not all studies consider the effects of the immersing liquid on the rotor motions.

Hydrodynamical forces transmitted from one particle to another through the viscous fluid are crucial in explaining certain dynamics observed in dense suspensions of swimming bacteria \cite{Lushi2014}, the most commonly studied type of active matter. Hence, hydrodynamic interactions may have a significant  impact on the collective motion of other types of active particles. Active systems consisting of particles rotating due to a magnetic field display macroscopic generated fluid flows \cite{Snezhko2015}. Previous theoretical studies \cite{Lushi2015} highlighted the critical role played by the hydrodynamic coupling of rotors in the phase behavior of the system. For example, a pair of opposite-spin spheres translates due to the particle mutual advection by the rotational flows generated by particle moving \cite{Leoni2010, Fily2012b}, whereas a few same-spin spheres co-orbit  around their center of mass\cite{Lushi2015}. In the absence of hydrodynamic interaction the rotors location will remain ``frozen'' in space \cite{NguyenKEG14}. Likewise, the phase behavior of denser rotor populations is also sensitive to the hydrodynamic interactions \cite{YeoLV15}.

Here we quantify the dynamical behavior (microstructure, clustering, as well as the transport and diffusion) of passive spheres immersed in an active rotor bath. We consider numerically the flow generated by a monolayer of rotors embedded in a fluid, a similar  set-up as in our previous study with only rotors \cite{YeoLV15}. We show that the emerging dynamics varies greatly with rotor and inert particle densities and that the particles (rotor and inert) transport and diffusion depend on the system composition.

\begin{figure}[htp]
  \centering
    \vspace{-0.1in}
  \includegraphics[width=0.45\textwidth]{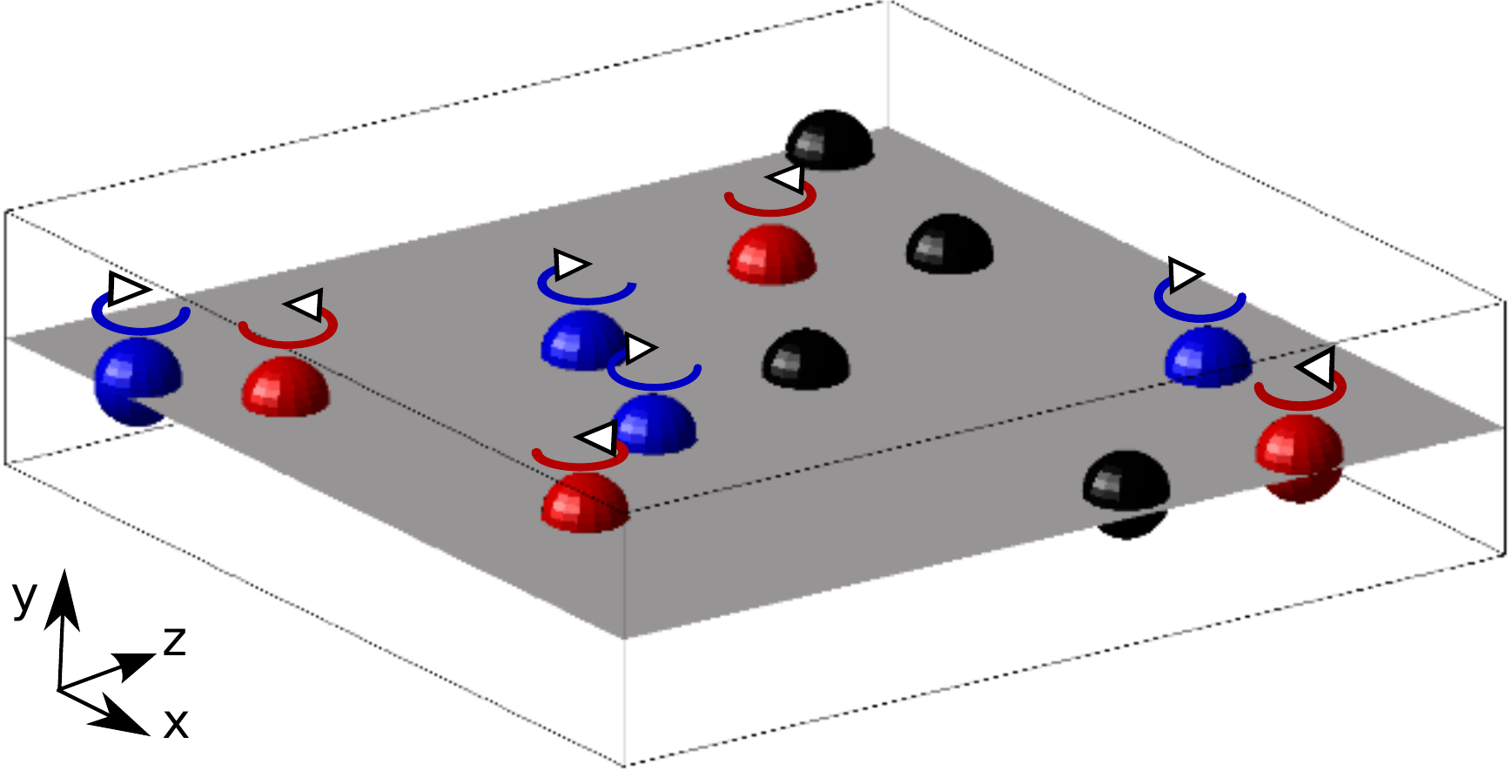}
    \vspace{-0.1in}
  \caption{  \footnotesize 
    Illustration of the system. The red- and blue-colored spheres denote the rotors whereas the black-colored spheres denote the inert particles. }
  \label{sketch}
  \vspace{-0.1in}
\end{figure}

%%%%%%%%%%%%%%%%%%%%%%%%%%%%%%%%%%
\section{Model}
\subsection{Particle motion}
We consider a mixture consisting of passive (inert) spheres  and rotating spheres (rotors) in the Stokes flow regime, where inertia effects are negligible. All the particles are placed in a monolayer, as illustrated in Figure \ref{sketch}.

An isolated  sphere with radius $a$  centered at  $\mathbf{x}_i$ and subjected to a constant torque $\tau$ generates a rotlet disturbance fluid flow   $\bm{u}_R(\mathbf{x}, \mathbf{x}_i)= \mathbf{\tau} \times (\mathbf{x} - \mathbf{x}_i) a^3/|\mathbf{x} - \mathbf{x}_i|^3$; its velocity decays slowly with the distance from the rotor as $\sim 1/r^2$. In a collection of rotors, the flow stirred by each  particle drags the other particles. The particles positions and rotations evolve as \cite{YeoLV15}
\begin{equation}
\label{eqM}
\begin{split}
\bm{V} = \frac{d\mathbf{x}_i}{dt}&= \sum_{j \neq i}\bm{u}_R(\mathbf{x}_i, \mathbf{x}_j) + \bm{u}_{corr}\\
\bm{\Omega}_i &= \bm{\Omega}_{0i} + \frac{1}{2} \sum_{j \neq i} \nabla \times  \bm{u}_R(\mathbf{x}_i, \mathbf{x}_j) + \bm{\Omega}_{corr}.
\end{split}
\end{equation}

Here $\bm{u}_{corr}$ is the correction fluid velocity that comes from multi-body interactions as well as lubrications. For pairwise interactions, the corrections to the fluid flow are $O(1/r^7)$ as calculated in \cite{YeoLV15}, but for multi-body interactions and in periodic domains these change \cite{KimKarrila}.
$\Omega_0=|\mathbf{\tau}|/8\pi \mu a^3$ is the rotation rate of an isolated rotor, where $\mu$ is viscosity of the embedding fluid. We neglect thermal noise in Eq.\ref{eqM} under the assumption of strong convection by the fluid flow \cite{YeoLV15}. In the equations above, for passive particles $\mathbf{\tau}_i = \mathbf{0}$, whereas for left/right-spinning particles $\mathbf{\tau}_i = \pm |\mathbf{\tau}| \mathbf{\hat{y}}$. In this study we take the magnitude of the applied torque $|\mathbf{\tau}|$ to be constant for all the rotor particles.

%%%%%%%%%%%%%%%%%%%%%%%%%%%%%%%%%%
\subsection{Numerical Method}

In dilute suspensions, the collectively-generated fluid flows can be approximated by a superposition of the appropriate rotlet flows. However, in dense suspensions where particles can get closer, the full hydrodynamic interactions with higher-order multipoles and the lubrication flows cannot be ignored. These interactions become too complicated to resolve analytically for more than a few particles \cite{Stickel2005}. 

Here, the full hydrodynamic interactions between the particles are computed using the force-coupling method (FCM)\cite{Yeo10e}. The coupled system of the long-range multi-body interactions is resolved with regularized low-order multipoles whereas the short-range lubrication interactions are approximated by the summation of pair-wise analytical solutions. The force-coupling method is well-suited to handle the dynamics of a large number of particles and has in the past been successfully applied to study various suspension flows \cite{Climent04,Yeo10a,YeoLV15}. 

We outline here the equations to be solved to obtain the fluid flow which is needed to resolve the particles' equations of motion. The fluid flow in this low Reynolds number regime is described by the following equations
\begin{align}
&\nabla p = \mu \nabla \cdot \nabla \bm{u} + \sum^{N_p}_{i=1} \left\{ \bm{F}_i \Delta_M(\bm{r}^n)+ \bm{G}_{i} : \nabla \Delta_D(\bm{r}^n)\right\},  \nonumber \\
&\bm{\nabla} \cdot \bm{u} = 0. \label{Eqn:Continuity}
\end{align}

Here, $p$ is pressure, $\bm{u}$ is fluid velocity, $\bm{r}_i$ is the position vector from a particle center ($\bm{r}_i = \bm{x}-\bm{x}_i$) , and $\bm{F}$ and $\bm{G}$ are the force monopole and force dipole moments, respectively. Here $( \bm{G}_{i} : \nabla \Delta_D(\bm{r}^i) )_k = \bm{G}_{kl}^i\frac{\partial}{\partial x_l}\Delta_D(\bm{r}^i)$.
The force envelopes $\Delta_M$ and $\Delta_D$ are given by
\begin{align}
\Delta_M(\bm{r}) &= \frac{1}{(2 \pi \sigma_M^2)^{3/2}} \exp{ \left( -\frac{\bm{r}^2}{2 \sigma_M^2} \right)}, \\
\Delta_D(\bm{r}) &= \frac{1}{(2 \pi \sigma_D^2)^{3/2}} \exp{ \left( -\frac{\bm{r}^2}{2 \sigma_D^2} \right)},
\end{align}
in which $\sigma_M = a/\sqrt{\pi}$ and $\sigma_D = a/(6\sqrt{\pi})^{1/3}$. The force monopole and dipole moments are
\begin{align}
\bm{F} = \bm{F}^S - \bm{F}^{lub}_i,  \  \   \  \  \   \ 
\bm{G} = \bm{S}^{FCM} - \bm{C}^{lub} + \bm{C}^S.
\end{align}
Here, $\bm{F}^S$ is the steric interaction force between the particles, $\bm{S}^{FCM}$ is the FCM stresslet, and $\bm{F}^{lub}$ and $\bm{C}^{lub}$ are the Stokeslet and couplet coefficients from the lubrication interaction, which are computed from pair-wise analytical solutions. $\bm{C}^S$ is the couplet from the applied (or intrinsic) torque of the spinners; $\bm{C}^S_{ij} = \frac{1}{2}\epsilon_{ijk}\bm{\tau}_k$.

Once $\bm{u}$ is computed by solving Eqs. \ref{Eqn:Continuity}, the particle translational velocity $\bm{V}$ and rotation rate $\bm{\Omega}$ are obtained by
\begin{align}
\bm{V}^n &= \int \bm{u}(x) \Delta_M(\bm{r}^n) d^3 \bm{r},  \\%\  \   \  \  \   \ 
\bm{\Omega}^n_i &= \frac{1}{2} \int \epsilon_{ijk} \frac{\partial u_k}{\partial x_j} \Delta_D(\bm{r}^n) d^3 \bm{r}.
\end{align}
Since $\bm{F}^{lub}$, $\bm{\Omega}^{lub}$, and $\bm{S}^{FCM}$ are functions of both $\bm{V}$ and $\bm{\Omega}$, an iterative procedure is necessary to solve the system \cite{Yeo10e}. Then the particle position and rotation are advanced by $\bm{V}$ and $\bm{\Omega}$.

To model the steric or excluded volume effects, we employ a contact force model. The contact force on particle $i$ from particle $j$ is given by
\begin{equation} \label{Contact_Force}
\bm{F}_{ij}^S = 
  - F_{ref} \left( \frac{R_{ref}^2-|\bm{d}|^2}{R_{ref}^2-4a^2} \right)^6 \frac{\bm{d}}{|\bm{d}|}  \text{if $|\bm{d}| < R_{ref}$} %\\
\end{equation}
where $\bm{d} = \bm{x}_i - \bm{x}_j$, $F_{ref}$ is a constant, and $R_{ref}$ is a cut-off distance. In this study, the steric interaction is activated when the shortest distance between two particle surfaces ($\epsilon$) is less than $0.002a$, {\it i.e.} $R_{ref}/a = 2.002$. $F_{ref}$ is chosen to keep the minimum separation distance $\epsilon_{min} \simeq 0.001a$. The typical time-step used is $10^{-3}$.

The numerical simulations of the monolayer suspensions are performed in a computational domain of $H_x \times H_y \times H_z = 80a \times 20a \times 80a$, in which $y$ is the direction in which torques are applied. Periodic boundary conditions are used in the $x$- and $z$-directions. The particle monolayer is located at $y = 0$ and the computational box is bounded by rigid walls located at $y = \pm H_y/2$. The vertical separation $20a$ is chosen big enough to guarantee that the wall boundaries do not affect the dynamics at the monolayer. Note that since the generated hydrodynamic flows due to rotation are azimuthal in nature and do not induce particle translation perpendicular to the monolayer, the particles remain within the monolayer.

We consider suspensions of passive spheres inter-dispersed in a bath of  rotors. The active phase consists of 50:50 mixture of opposite-spin rotors; clockwise- and counterclockwise-spinning spheres of equal number. The total volume fraction of all the particles (passive and spinning) varies from $\phi = 0.06$ to $0.46$. The volume fraction of passive particles varies from $\phi_P=0.02$ to $0.38$, while the volume fractions of the rotors are $\phi_R = 0.04$, 0.08, 0.16, 0.20, and 0.24. For a monolayer suspension, the volume fraction is defined as $\phi = (\frac{4}{3}\pi a^3) N_p /  (H_x \times H_z \times 2a)$, in which $N_p$ is the number of the particles.
The magnitude of the external torque is determined to make the reference angular velocity of the spinners be $\Omega_0= \pm 1$. All of the simulations start from initial random configurations, generated by a molecular dynamics procedure. The dynamics are studied after the suspensions reach stationary states, typically about $t  \simeq O(10^4)$ from the initial time (time is non-dimensionalized by $\Omega_0$).

Note that the all the particles, inert and spinning, are interacting hydrodynamically with each-other due to singularities of higher order than rotlets as well as short-range lubrication flows, denoted as $\bm{u}_{corr}$ in Eqn. \ref{eqM}. The total fluid flows are not just a superposition of rotlets, and the passive spheres do contribute to the collectively generated flows by adding short-ranged fluid disturbances.

%%%%%%%%%%%%%%%%%%%%%%%%%%%%%%%%%%

\section{ Structuring and phase diagram}
Here we present the simulations results for particle self-organization. Physical interpretation and analysis of the observed behaviors are provided in Section \ref{sec:Results}.

\begin{figure}[htp]
  \centering
   \includegraphics[width=0.48\textwidth]{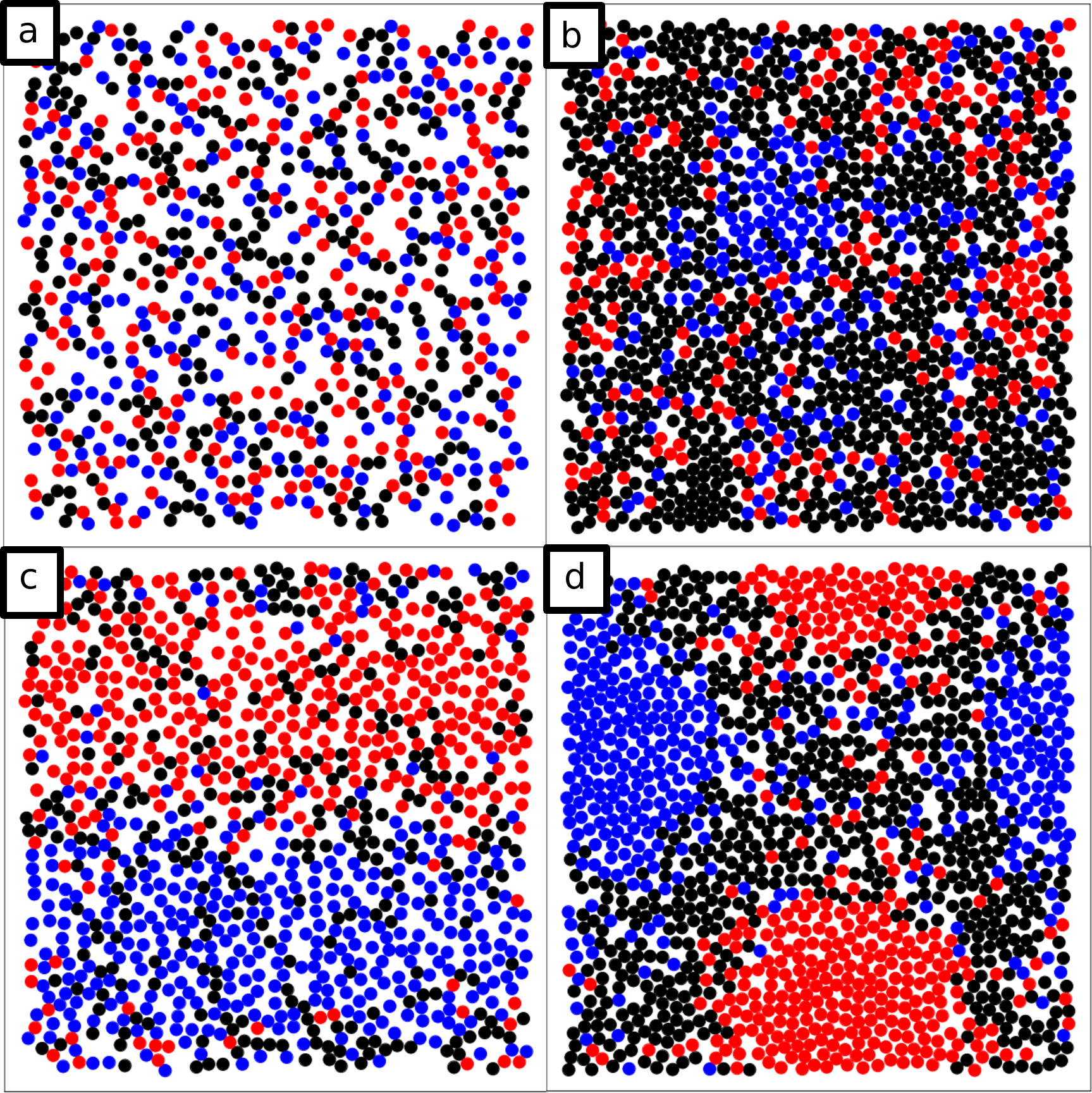}
   \vspace{-0.1in}
  \caption{  \footnotesize 
    Snapshots of the system seen from above at long times for (a) $\phi_R = 0.16,\phi_P = 0.1$, (b) $\phi_R=0.16,\phi_P=0.3$, (c) $\phi_R=0.24,\phi_P=0.1$, and (d) $\phi_R=0.24,\phi_P=0.2$. The red/blue-colored spheres denote the left/right-spinning rotors whereas the black-colored spheres denote the passive particles. }
   % \vspace{-0.1in}
  \label{snapshots}
\end{figure}

Snapshots of the system for rotor densities $\phi_R = 0.16$,$0.24$ and various passive particle densities are shown in Figure \ref{snapshots}. 

In our previous study on binary rotor mixtures\cite{YeoLV15}, we demonstrated that the suspensions of active rotors remain in a gas-like state at low total rotor volume fractions ($\phi_R \le 0.2$). With increased rotor density  rotors of different spin segregate forming phase-separated fluid phases (macroscopic lanes or vortices).

Similarly here, at $\phi_R = 0.16,\phi_P = 0.1$ shown in Figure \ref{snapshots}a, the system is in a gas-like state, which changes to a phase-separated fluid state at increased rotor density $\phi_R = 0.24, \phi_P = 0.1$, as seen in Figure \ref{snapshots}c. When the rotor density is fixed at $\phi_R = 0.16$ but tracer density $\phi_P$ is increased from $0.1$ to $0.3$, we find that the rotors form clusters of same-spin rotors, see Figure \ref{snapshots}b. At $\phi_R = 0.24, \phi_P = 0.1$ the rotors segregate into fluid regions made of same spin rotors whereas the passive particles can be found in either phase.

When the passive particle density $\phi_P$ is increased to $0.2$, as seen in Figure \ref{snapshots}d, two counter-rotating vortices emerge and the passive particles are now moving along the boundaries of the phase-separated fluids. Compared to our previous study with only rotors \cite{YeoLV15} where the counter-rotating vortices are observed when the rotor density reaches $\phi_R = 0.5$, here the phase transition occurs at a much lower rotor density $\phi_R$ and lower total particle density $\phi_{R+P}=0.44$. This effect is  due to the presence and excluded volume of the passive particles.

\begin{figure}[htp]
  \centering
  \includegraphics[width=0.40\textwidth]{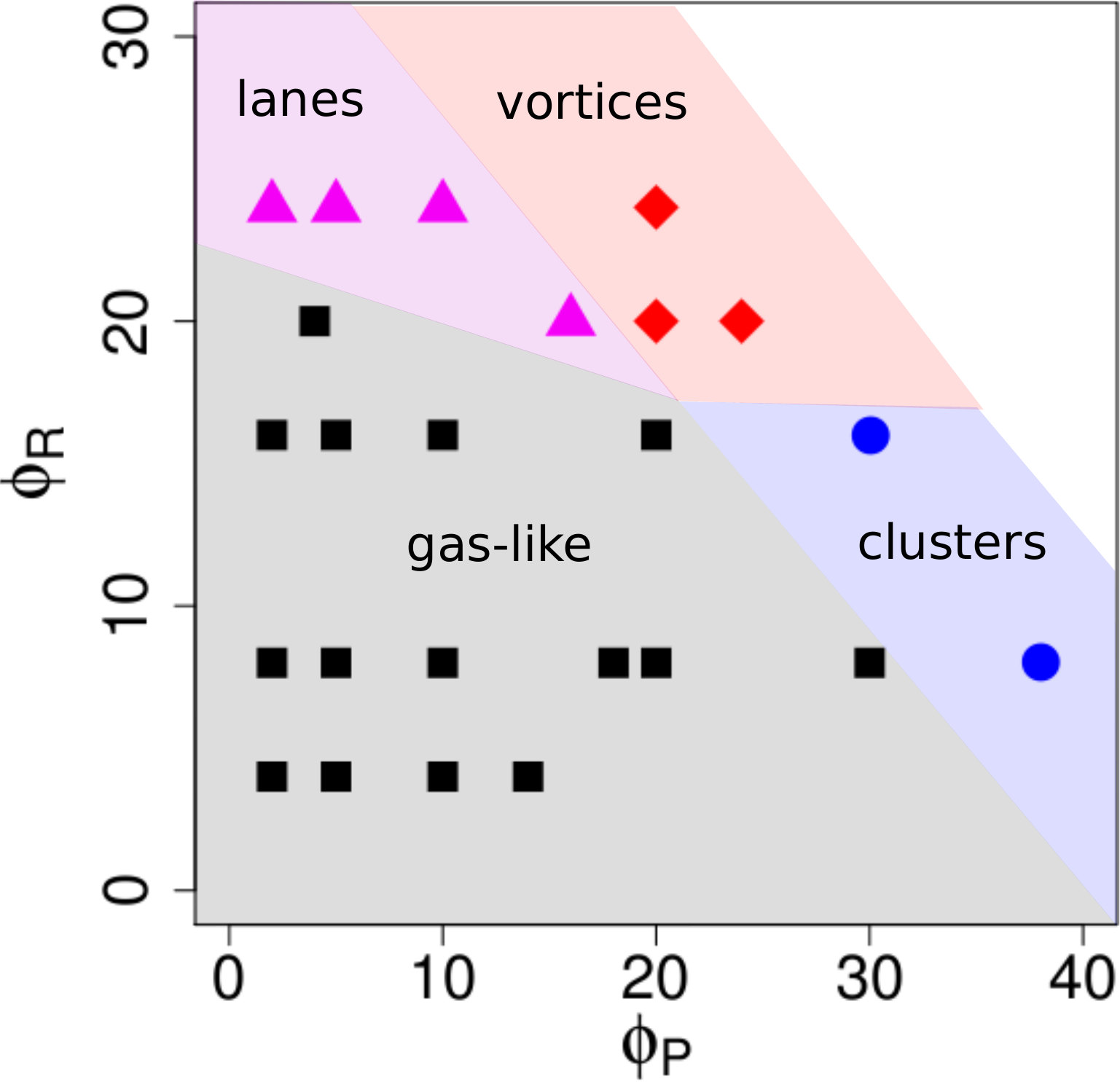}
      \vspace{-0.1in}
  \caption{  \footnotesize Phase diagram of various dynamical states obtained by the system as a function of the rotor density $\phi_R$ and passive tracer density $\phi_P$. The symbols represent the various states: $\blacksquare$ for a dilute gas-like phase, 
  ${\color{blue}\bullet}$ for the state with clustering of rotors and clustering of tracers, 
  ${\color{magenta}\blacktriangle}$ for the state where the rotors have phase-separated but the tracers are interdispersed throughout, 
  ${\color{red}\blacklozenge}$ for the state where the rotors self-organize into large vortices and crystals while the tracers move along the boundaries of the phase-separated rotor-fluid regions.% The shaded regions in the phase-space diagram are approximate.
    }\vspace{-0.1in}
  \label{phasediag}
\end{figure}

The dynamics observed at various rotor and inert-particle densities is summarized in the phase diagram in Figure \ref{phasediag}. The phase diagram is computed based on the analysis of microstructures shown in section \ref{subset:microstructure}.
\begin{itemize}
\item At low rotor and tracer densities, the prevailing dynamics is gas-like with rotors and tracers moving chaotically in the domain ($\blacksquare$), as in the example of Figure \ref{snapshots}a. 
\item At intermediate rotor densities but high tracer densities we observe clustering of the same-spin rotors (${\color{blue}\bullet}$), as in the example of Figure \ref{snapshots}b. 
\item At high rotor densities but low tracer densities, the rotors phase-separate whereas the tracers are scattered throughout the domain (${\color{magenta}\blacktriangle}$), as in the example of Figure \ref{snapshots}c. 
\item At high rotor and tracer densities, rotors self-assemble into large rotating crystals and the passive particles move along the boundaries of the rotating crystals (${\color{red}\blacklozenge}$), as in the example of Figure \ref{snapshots}d.
\end{itemize}

%%%%%%%%%%%%%%%%%%%%%%%%%%%%%%%%%%
\section{Discussion of the results} \label{sec:Results}

\subsection{Microstructures} \label{subset:microstructure}

To quantitatively study the phase transition and clustering of the rotors and tracers, we utilize the partial number density of particles of type $B$ around the reference particle type $A$ defined as \cite{YeoLV15}
\[
\lambda_{AB}(r) = \frac{1}{n_B}\left\langle \frac{ \sum_{j=1}^{N-1} H(r - |\bm{d}_{j}|)\chi_{AB}(j)}{\pi r^2 (2a)}   \right\rangle.
\]
Here $H(x)$ is the Heaviside function, $N$ is the number of the particles (rotors and passive), $|\bm{d}_{j}|$ is the distance from the reference particle, and $n_B$ is the number density of the $B$-type particles. The indicator function $\chi_{AB}(j)$ is one when the relation between the reference and $j$-the particles satisfies the set $AB$, and zero otherwise. Note that $\lambda_{AB}(r) \rightarrow 1$ for large $r$.

First we look at the clustering of rotors by computing the partial density for rotors of the same spin $\lambda_{++}$ and the partial density of rotors of the opposite spin $\lambda_{+-}$.

\begin{figure}[htps]
  \centering
    \includegraphics[width=0.42\textwidth]{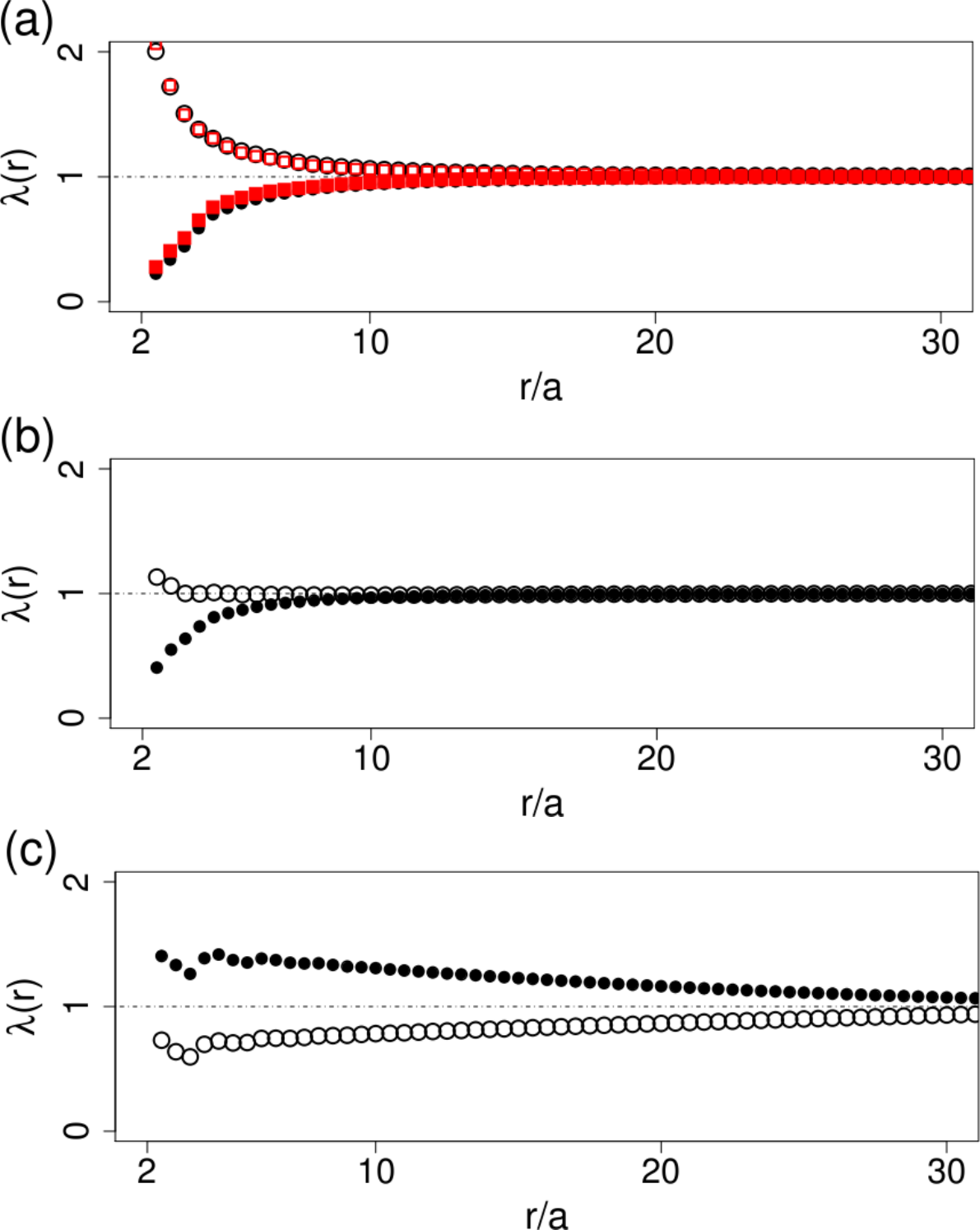}
  \caption{  \footnotesize 
     Rotor partial densities $\lambda_{++}$ ($\bullet$) and $\lambda_{+-}$ ($\circ$) for rotor and tracer densities (a) $\phi_R=0.08,\phi_P=0.1$(in black), $\phi_P=0.2$ (in red), (b) $\phi_R=0.16,\phi_P=0.1$, and (c) $\phi_R=0.16,\phi_P=0.3$.  }
     \vspace{-0.1in}
  \label{lambda}
\end{figure}

Figure \ref{lambda}a shows the partial densities for same-spin rotors $\lambda_{++}$ and opposite-spin rotors $\lambda_{+-}$ for low rotor density ($\phi_R=0.08$) at two passive particle densities $\phi_P=0.1,~0.2$. We observe that $\lambda_{+-} \ge \lambda_{++}$ for small $r$, which implies that the opposite spin rotors tend to move together, similar to our previous results without the passive particles \cite{YeoLV15}. Interestingly, at low rotor density, increasing the passive particle density $\phi_P$ from $0.1$ (black marker) to $0.2$ (red marker) does not affect $\lambda_{++}$ or $\lambda_{+-}$.

As the rotor density increases to $\phi_R=0.16$, the pairing of opposite spin rotors weakens (figure \ref{lambda}b). However, still $\lambda_{+-}$ is larger than $\lambda_{++}$.

As shown in Figure \ref{lambda}c, increasing the passive particle density $\phi_P$, while keeping the rotor density fixed at an intermediate value ($\phi_R=0.16$), results in microstructural changes, from the pairs of the opposite spin rotors co-translating ($\lambda_{+-} > \lambda_{++}$) at $\phi_P = 0.1$ to clusters of the same spin rotors co-rotating ($\lambda_{+-} < \lambda_{++}$) at $\phi_P = 0.3$. Note that due to the hydrodynamic interactions, rotors of opposite spins tend to co-translate and rotors of same spin co-rotate \cite{Lushi2015}.

These results show that the addition of inert particles to an active rotor bath modifies the rotor-rotor correlations in non-trivial ways. At very low rotor densities, adding more inert particles does not seem to affect the tendency of opposite-spin rotors to co-translate. At slightly higher rotor density however, adding more tracers to the mixture can switch the tendency to rotors from opposite-spin pairings toward same-spin pairings.  

\begin{figure*}[htps]
  \centering
   \includegraphics[width=1.75\columnwidth]{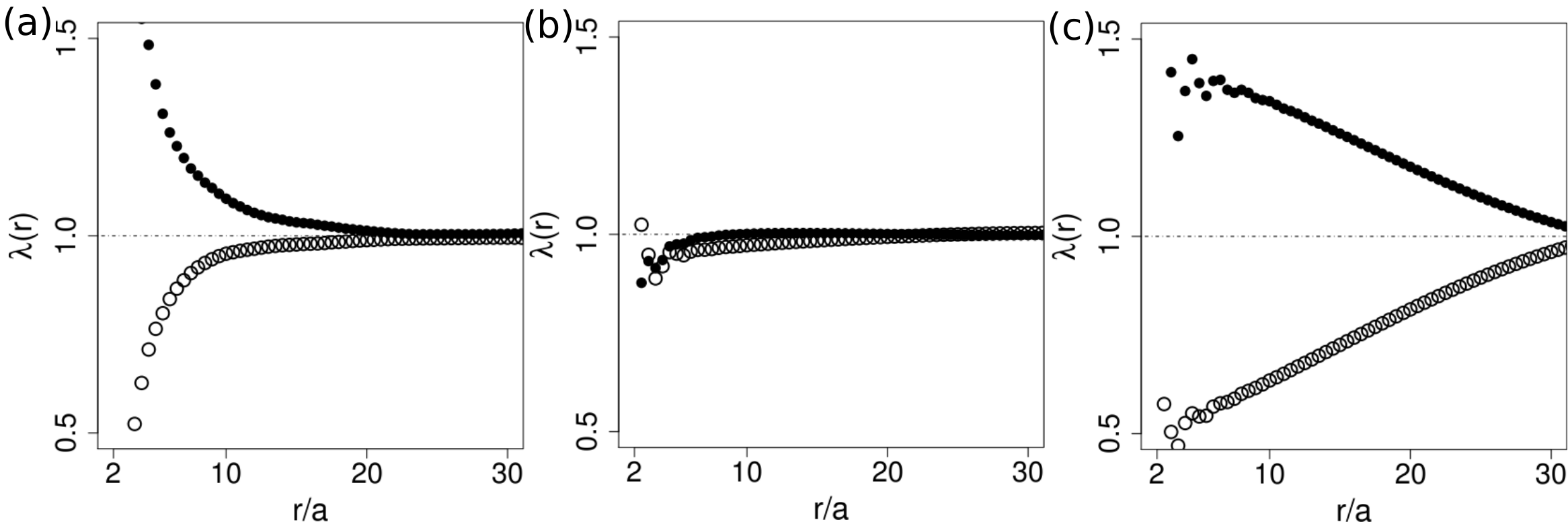}
   \vspace{-0.1in}
  \caption{  \footnotesize 
    Partial densities for rotor-rotor $\lambda_{RR}$ ($\bullet$) and rotor-tracer $\lambda_{RP}$ ($\circ$) in the cases (a) $\phi_R=0.04,\phi_P=0.1$, (b) $\phi_R=0.24,\phi_P=0.1$, and (c) $\phi_R=0.24,\phi_P=0.2$.}
  \label{pdfs}
\end{figure*}

%%%%%%%%%%%%%%%%%%%%%%%
\subsection{Rotor and inert particle correlations}

Now we try to quantify the clustering and correlations of the rotors and the inert particles by computing the rotor-rotor particle partial densities $\lambda_{RR}$ and rotor-passive particle partial densities $\lambda_{RP}$. Here we do not differentiate the rotors by spin.

Figure \ref{pdfs}ab displays the partial densities for the rotor-rotor ($\lambda_{RR}$) and the rotor-passive particles ($\lambda_{RP}$) for the same tracer density ($\phi_P=0.1$) but different rotor densities ($\phi_R = 0.04,~0.24$). At the lower rotor densities, in Figure \ref{pdfs}a we observe $\lambda_{RR} > \lambda_{RP}$, indicating that the rotors in general have a higher probability to cluster together. Increasing the rotor density %, where recall that translational kinetic energy of the tracers is higher than that of rotors $T_P > T_R$, 
$\lambda_{RR}\approx \lambda_{RP}$, as seen in Figure \ref{pdfs}b. However, when the passive particle density is also increased further, clustering or phase separation occurs and $\lambda_{RR}>\lambda_{RP}$ again, as seen in Figure \ref{pdfs}c. 

It should be noted that, although $\lambda_{RR} > \lambda_{RP}$ is observed both at the lower (Figure \ref{pdfs}a) and higher rotor densities $\phi_R$ (Figure \ref{pdfs}c), the micro-structural origins in these cases are different. As explained in the previous sections (Figures \ref{snapshots}, \ref{lambda}), the larger $\lambda_{RR}$ at low rotor density ($\phi_R=0.04$) is due to the formation of the doublets of opposite spin rotors co-translating, whereas at high rotor density  ($\phi_R=0.24$) the co-rotation and clustering of same spin rotors, or phase separation, is responsible for the larger $\lambda_{RR}$.

%%%%%%%%%%%%%%%%%%%%%%%%%%%%%%%%%%%
\subsection{Particle transport and mixing dynamics}

Here, we investigate the rotor and inert particles' transport and diffusion by analyzing the mean square displacements. 
\begin{figure}[h]
  \centering
    \includegraphics[width=0.47\textwidth]{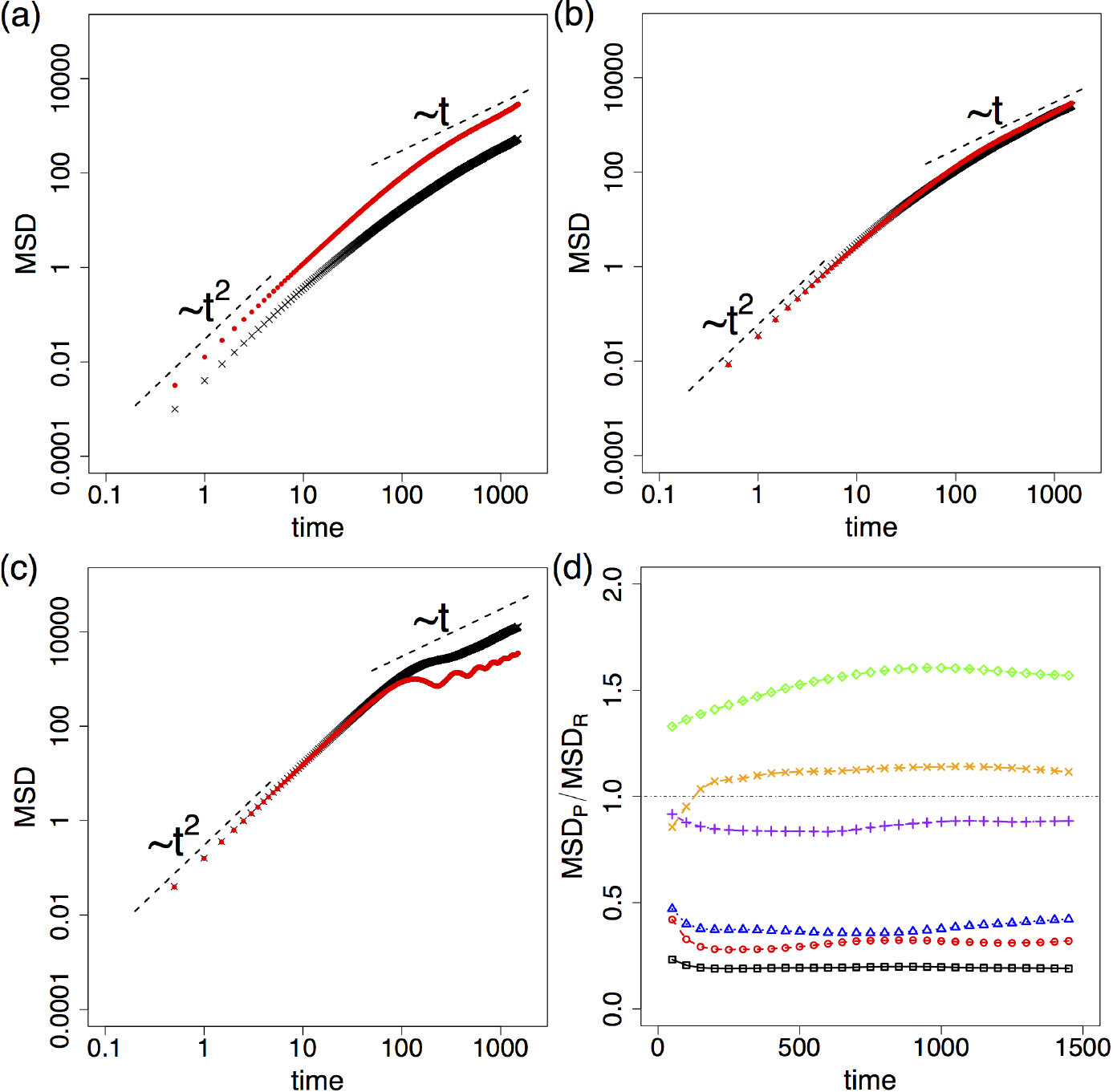}
    %    \vspace{-0.1in}
  \caption{  \footnotesize 
    The mean square displacements of the passive particles ($\times$) and the rotors (${\color{red}\bullet}$) in time for (a) $\phi_R=0.04$, $\phi_P=0.1$, (b) $\phi_R=0.16$, $\phi_P=0.1$, and (c) $\phi_R=0.24$, $\phi_P=0.2$. (d) The ratio of the MSD of the passive particles to the MSD of the rotors for $\phi_R = 0.04$,  $\phi_P=0.1$ denoted with $\square$, $\phi_R = 0.08$, $\phi_P=0.1$ denoted with ${\color{red}\circ}$, $\phi_R = 0.08$, $\phi_P=0.2$ denoted with ${\color{blue}\triangle}$, $\phi_R = 0.16$, $\phi_P=0.2$ denoted with ${\color{brightube}+}$, $\phi_R = 0.16$, $\phi_P=0.3$ denoted with ${\color{orange}\times}$, and $\phi_R = 0.24$, $\phi_P=0.1$ denoted with ${\color{green}\Diamond}$. }
    \vspace{-0.1in}
  \label{msd}
\end{figure}

Figure \ref{msd}a--c shows the mean-square displacements of the rotors ($\text{MSD}_R$) and the passive particles ($\text{MSD}_P$). At low rotor density (Figure \ref{msd}a), $\text{MSD}_R>\text{MSD}_P$, and rotors transport further than passive particles, but, as the rotor density increases, $\text{MSD}_R$ becomes similar to $\text{MSD}_P$ (Figure \ref{msd}b). 

Figure \ref{msd}c shows MSDs at the states where the rotors are self-organized into two counter-rotating vortices. In this phase-separated regime the passive particles move much longer distance than the rotors, mainly because the rotors exhibit spiral motion while trapped in the vortical structures, evident even in the oscillatory patten of $\text{MSD}_R$. 

In Figure \ref{msd}d we present the ratio of $\text{MSD}_P$ to $\text{MSD}_R$. It is clearly shown that at the low rotor densities ($\square$,${\color{red}\circ}$,${\color{blue}\triangle}$) $\text{MSD}_P \ll \text{MSD}_R$. At these low densities, the rotors have a high probability to form a doublet of opposite spin rotors which exhibits translational motion. Considering the collective translational motion of the rotors at low rotor density, it is not surprising to see that $\text{MSD}_R > \text{MSD}_P$ indicating rotors transport much farther than passive particles. 

At increased rotor density the microstructure of the rotor mixture changes from pairs of the opposite spin rotors at low tracer density  (${\color{brightube}+}$) to clusters of the same spin rotors at higher tracer density (${\color{orange}\times}$), as explained in the previous subsection. Following the changes in the microstructures of the rotors, the ratio $\text{MSD}_P / \text{MSD}_R$, which is less than one for low tracer density, becomes larger than one when the tracer density is increased, indicating increased transport of the passive particles. 

At high rotor density and low tracer density (${\color{green}\Diamond}$), there is a complete phase separation of the rotors and $\text{MSD}_P \gg \text{MSD}_R$. When the clusters of same spin rotors emerge, the rotors exhibit closed spiral motion in a cluster, generating a jet-like flow in the interstitial regions. The passive particles translate very long distances as they are carried by these jet-like flows in the boundaries of the clusters, whereas the rotors are trapped in the macroscopic vortices.

%%%%%%%%%%
\subsection{Particle motion and system energy}

In active suspensions, the suspended particles, both rotors and passive particles, exhibit translational as well as rotational motions in response to the flow generated by the active rotors. In other words, the system is driven by the rotational kinetic energy, which is subsequently converted to the translational kinetic energy through the hydrodynamic interactions. We quantify the effects on the systems by tracing the translational and rotational kinetic energies, defined respectively as
\[
T_A = \frac{1}{2} m \langle \bm{V}\cdot\bm{V} \rangle_A,  \  \   \  \  \   \ R_A = \frac{1}{2} I \langle \bm{\Omega}\cdot\bm{\Omega} \rangle_A,
\]
where $m$ is the mass of a particle, $I$ is the moment of inertia, and $\langle \cdot \rangle_A$ is an ensemble average over the active or passive particles. We also define the ratio
\[
\kappa_A = \frac{T_A}{R_A + T_A}.
\]

\begin{figure}[htp]
  \centering
  %\vspace{-0.1in}
   \includegraphics[width=0.5\textwidth]{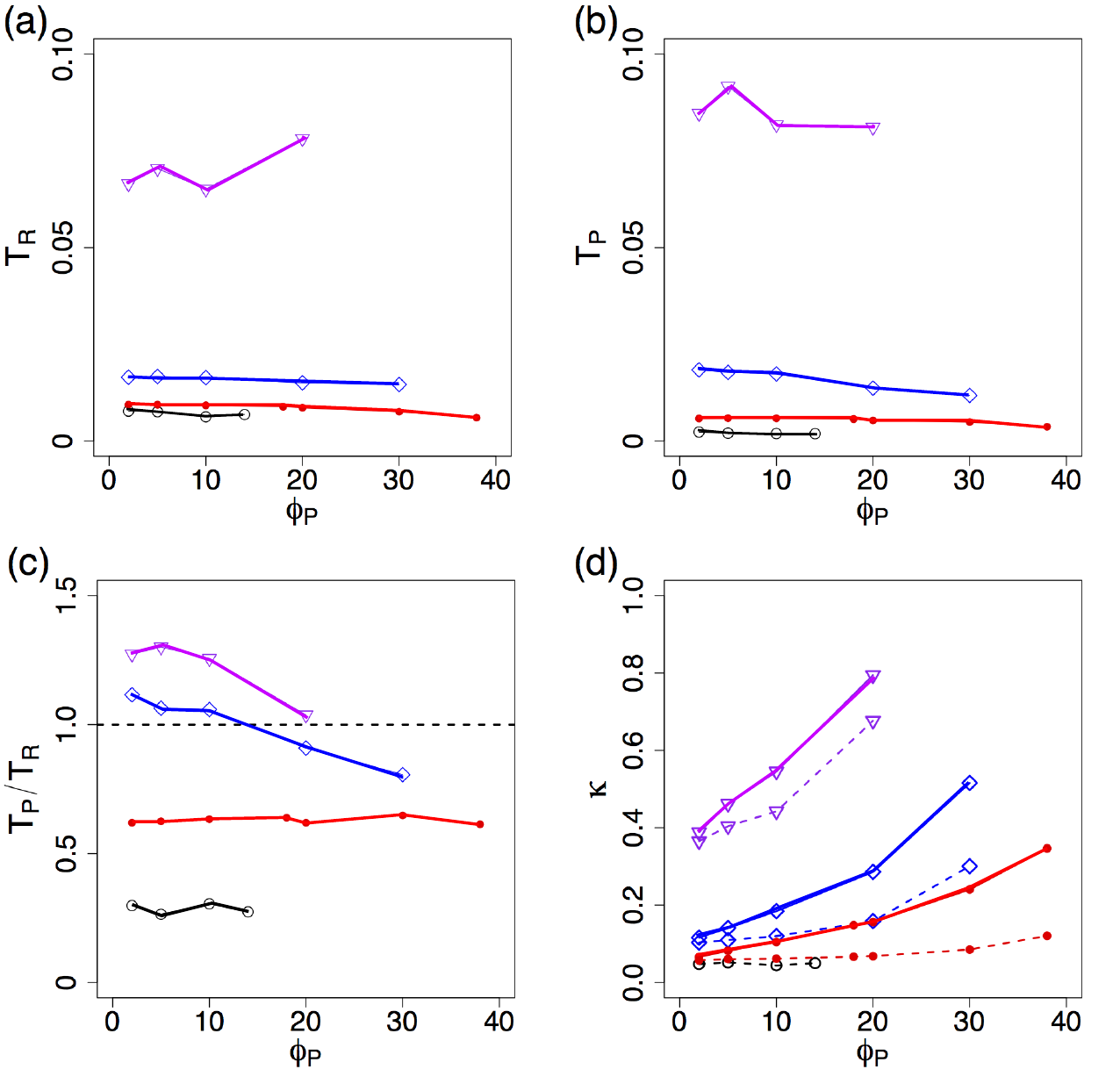}
    %\vspace{-0.2in}
  \caption{  \footnotesize 
    (a) Translational kinetic energy of the rotors $T_R$ and (b) of the passive particles $T_P$ as functions of passive particle density $\phi_P$ for four different rotor densities: $\circ$: $\phi_R = 0.04$, ${\color{red}\bullet}$: $\phi_R = 0.08$, ${\color{blue}\diamond}$: $\phi_R=0.16$, and ${\color{magenta}\nabla}$: $\phi_R = 0.24$. (c) The ratio of the translational kinetic energies of the passive particles to that of rotors  $T_P/T_R$. (d) The ratio of the translational kinetic energies to the total kinetic energy of the rotors (dashed line) and the entire system (solid line).  }
  \label{fig1}
  % \vspace{-0.1in}
\end{figure}

Figure \ref{fig1}a shows changes of the rotors' translational kinetic energy $T_R$ as a function of the passive particle density $\phi_P$ for four different rotor densities. At low rotor densities ($\phi_R \le 0.16$), $T_R$ does not change significantly with increasing passive particle density. For example, $T_R$ for $\phi_R = 0.16$ is reduced only about 12\% when $\phi_P$ changes from $0.02$ to $0.3$. Intuitively, for a fixed rotor density, the translational kinetic energy is expected to be a decreasing function of tracer density since the viscous dissipation increases with increasing tracer density. However, at high rotor density $T_R$ becomes a non-monotonic increasing function of tracer density. %Indeed, $T_R$ increases for $\phi_P \ge 0.1$.

Figure \ref{fig1}b shows the translational kinetic energy of the passive particles $T_P$ as a function of the density of passive particles $\phi_P$ for four different rotor densities. For a fixed tracer density, $T_P$ is a more sensitive function of rotor density than $T_R$. For example, at $\phi_P = 0.05$, increasing rotor density $\phi_R = 0.04\rightarrow 0.24$ increases $T_P$ 46-fold but increases $T_R$ only 9-fold. $T_P$ shows a behavior similar to $T_R$ at low rotor densities: a decrease of $T_P$ at higher passive particle density $\phi_P$.

The ratio of the tracer and rotor translational kinetic energies $T_P/T_R$ is presented in Figure \ref{fig1}c. For the two large rotor densities and small tracer densities, $T_P/T_R>1$, indicating that the passive particles move faster than the active rotors. However, $T_P/T_R$ decreases with tracer density and becomes less than one for $\phi_P > 0.20$. On the other hand, for the two low rotor densities $T_P/T_R$ is almost uniform. Particularly, for very low rotor density $\phi_R = 0.08$ this ratio kinetic energies does not change significantly for a wide range of tracer density.

In Figure \ref{fig1}d we plot the ratio  $\kappa$ of the translational to the total (translational + rotational) kinetic energy of the rotors. 
Similar to previous studies \cite{NguyenKEG14,YeoLV15}, $\kappa$ is an increasing function of the rotor density when the passive particle density is held constant. At very low rotor densities $\kappa$ is relatively uniform for the range of $\phi_P$ in the present study, whereas for increased rotor density $\kappa$ increases fast with larger $\phi_P$.  

Figure \ref{fig1}d also shows $\kappa$ of the entire system (passive + active particles) 
\[
\kappa^{total} = \frac{\phi_R T_R + \phi_P T_P}{\phi_R(T_R+R_R)+\phi_P(T_P+R_P)}.
\]
Notice that $\kappa^{total}$ becomes larger than $\kappa$ of the rotors. It is found that the rotational kinetic energy of the passive particles ($R_P$) is negligible compared to $T_P$. In other words, $\kappa_P \simeq 1$. As a result, adding passive particles to the active rotor system makes the ratio of the translational kinetic energy to the total kinetic energy larger than that of the active rotor system. \\

%%%%%%%%%%%%
\section{Conclusion and outlook}

We have numerically investigated the collective dynamics in a monolayer mixture of inert and rotating spheres immersed in a fluid.  We find that passive particles modify the phase behavior of the 50:50 mixture of opposite spin rotors and the phase boundaries are shifted. Although the structuring shows qualitatively similar behavior to the purely active system \cite{YeoLV15} (gas-like phase, spin-separated lanes or vortices), here they occur at different particle densities. 

The system microstructure, particle transport and diffusion are affected in non-trivial ways by the addition of the tracers. In intermediate rotor densities, the addition of inert particles can switch the tendency of the rotors to group with same-spin rotors instead of grouping with opposite-spin ones. At low rotor and tracer densities, the rotors travel further than the tracers due to opposite-spin rotors co-translating. At high rotor densities, rotors spin-separate into large clusters or vortices due to same-spin rotors co-rotating; rotors then are trapped in the macroscopic clusters, whereas inert particles get transported far by the fluid flows generated in the interstitial regions between the rotor clusters. 

Adding inert particles in an active micro-rotor bath yields unexpected outcomes that are delicately dependent on the particle densities and ratios. This suggests that the dynamics of passive particles in other types of active suspensions, a topic of growing interest \cite{Kim2004, Mino2011, Angelani2011, Jepson2013, Kasyap2014, Patteson2015, Cheng2015}, is non-trivial, and that transport and diffusion of the particles can be affected in unforeseeable  ways.

We anticipate growing interest in the topic of micro-rotor given the increase of practical realizations of self-rotating particles in synthetic systems, e.g., applying a torque to a particle by magnetic \cite{Grzybowski2000, Grzybowski2002, Kokot2013, Kokot2015, Snezhko2015}, electrical \cite{Bricard2013, Bricard2015} or optical fields \cite{Friese1998}, or biological systems, e.g.,  {\it T. majus} bacteria \cite{Petroff2015},  {\it Volvox} algae \cite{Drescher2009} or aggregates of swimming bacteria  \cite{SchwarzLinek2012}. We hope our numerical work will provide useful insights to understand the behavior of existing experimental systems and design new ones. 

\vspace{0.3in}

\section*{Acknowledgements}
E. Lushi and P.M. Vlahovska acknowledge support from the NSF through awards CBET 1437545 and CBET 1544196.

%%%%%%%%%%
%\begin{thebibliography}{99}

\end{document}